# Broadband transmissive polarization rotator by gradiently-twisted α-MoO$_3$

Songyan Hou,[1,2,a)] Hao Hu,[3,a)] Zhihong Liu,[1,2)] Weichuan Xing,[1,2)] and Jincheng Zhang,[1,2,a)]

[1] *Guangzhou Institute of Technology, Xidian University, Guangzhou 510555, China*

[2] *State Key Laboratory of Wide Bandgap Semiconductor Devices and Integrated Technology, School of Microelectronics, Xidian University, Xi'an 710071, China*

[3] *Key Laboratory of Radar Imaging and Microwave Photonics Ministry of Education, College of Electronic and Information Engineering, Nanjing University of Aeronautics and Astronautics, Nanjing 211106, China*

Authors to whom correspondence should be addressed: Songyan Hou, housy1@outlook.com; Hao Hu, hao.hu@nuaa.edu.cn; and Jincheng Zhang, jchzhang@xidian.edu.cn

Polarization engineering has been proven to enhance the capabilities of light manipulation and thus facilitate the development of integrated photonic devices. In this study, we introduce a polarization rotator based on gradiently-twisted α-MoO$_3$ thin film, that works for the mid infrared range and functions in a transmission mode. To be specific, the proposed device is constructed by gradiently-twisted α-MoO$_3$ multilayers with a subwavelength thickness of only 5 microns, namely, one-third of the working wavelength. Our analytical calculation demonstrates the efficacy of this subwavelength thin film rotator in converting a linearly polarized wave into its orthogonal counterpart, thanks to its chiral nature. The twisted α-MoO$_3$ multilayers exhibit the capability to significantly manipulate dispersion characteristics while maintaining low optical losses, thereby enabling a wide bandwidth exceeding 2.5 THz with a polarization ratio surpassing 17 dB. Moreover, the operational frequency can be adjusted across a 3.4 THz range by altering the incident angle of the incident waves. This adaptable design, characterized by its polarization versatility, can be customized to suit practical applications within wireless communication, radar systems, optical switching, and imaging technologies.

Polarization control serves as a fundamental mechanism that facilitates various practical applications, including high-contrast imaging[1-4], wireless communications[5-8], and sensing[9-11]. It is well-known that coherent light sources are typically polarized in a specific direction. The polarization of light can be treated as a degree of freedom for the information processing, underscoring the importance of polarization rotators in these contexts. In the traditional approach, the modulation of the polarization direction of a linearly polarized wave is achieved using birefringent crystals[12] such as liquid crystals[13]. However, these methods typically require a relatively thick medium with a narrow-band performance. Recently, a broadband polarization converter based on metamaterials has been proposed and demonstrated, showcasing a remarkable capability to rotate a linear polarization state into its orthogonal counterpart[14, 15]. These metamaterials consist of a periodic array of subwavelength resonant structures, which can effectively control the phase of an incident wave, resulting in wideband polarization conversion. Additionally, their low profile, typically less than the operational wavelength, makes them less lossy and relatively straightforward to be fabricated using conventional semiconductor micro-fabrication methods. Despite these accomplishments, a common limitation of such devices is that the working frequency is determined by its design and cannot be tuned in a large range. Moreover, these metamaterial-based optical rotators work with reflection mode only, while the transmissive polarization rotators are more user-friendly compared to reflective counterparts. From a practical standpoint, transmissive polarization controllers assume paramount importance, given the restricted feasibility of applications associated with reflective counterparts. They are more straightforward to be integrated into optical setups and have advantages in terms of alignment and overall system complexity[16].

Recently, stacked hyperbolic crystals with twisted angle, so called twisted hyperbolic crystals, are shown with dramatic dispersion tunability[17-19]. In twistronics, when two-dimensional (2D) materials are twisted at a small angle, i.e., the magic angle, the moire superlattice is formed which exhibits exotic electrical properties such as unconventional flat-band superconductivity. The giant optical anisotropy has been observed in some van der Waals crystals, such as $MoO_3$[20, 21], $V_2O_3$[22] and $WTe_2$[23], which naturally induces a large intrinsic birefringence. Inspired from twistronics, the twisted hyperbolic crystals arise if the 2D materials are replaced by hyperbolic van der Waals crystals. The twisted hyperbolic crystals are highly interesting because they can enable the propagation of moire hyperbolic plasmons[24-26]. The isofrequency contour of moire hyperbolic plasmons exhibits rich topologies at quite large twisted angles[27], allowing for the flexibly control of near-field light flow at the mesoscopic scale.

In addition to the near-field control of light, the twisted hyperbolic crystals provide an enticing platform to engineer the light in the far field[28, 29]. This is intuitive because the rotational symmetry of twisted hyperbolic crystals is broken. As such, the twisted hyperbolic crystals can be effectively treated as a chiral medium that preserves the optical activity[30]. In other words, the twisted hyperbolic crystals offer a promising platform to control the polarization direction of incidence. Recent advances have already demonstrated the optical activity in a double-layer twisted hyperbolic metasurface[17]. Compared to conventional approaches, the twisted hyperbolic crystals are compact, broadband and reconfigurable[29]. However, due to the limited number of stacked layers, the previously realized optical activity is relatively weak. Up to date, how to achieve a high-performance polarization rotator based on twisted hyperbolic crystals still remains a scientific challenge.

As a natural van der Waals crystal, α-phase molybdenum trioxide (α-$MoO_3$) has been widely investigated due to its novel optical dispersion properties[21]. In this paper, we present gradiently-twisted α-$MoO_3$ polarization rotator to achieve broadband linear polarization control. The subwavelength device features a total thickness less than one-third of the central wavelength within the operational bandwidth. The chiral nature of gradiently-twisted α-$MoO_3$ leads to the polarization rotation, which has been demonstrated via the analytical calculation. The proposed device demonstrates a wide bandwidth of 2.5 THz with a polarization ratio over 17 dB and a conversion efficiency approaching unity. Remarkably, the operational frequency can be tuned over a range of 3.4 THz by adjusting the incident angle of the incoming waves. Moreover, the rotators with subwavelength thickness are lossless and fabrication-friendly compared to integrated rotators with complicated fabrication processes[31]. Our studies contribute an opportunity to realize high-performance subwavelength optical polarization rotator and the features of proposed rotator also facilitate the practical applications of integrated nanophotonics[32-39], biosensing[40], wireless communications[41] and imaging[1, 2, 4].

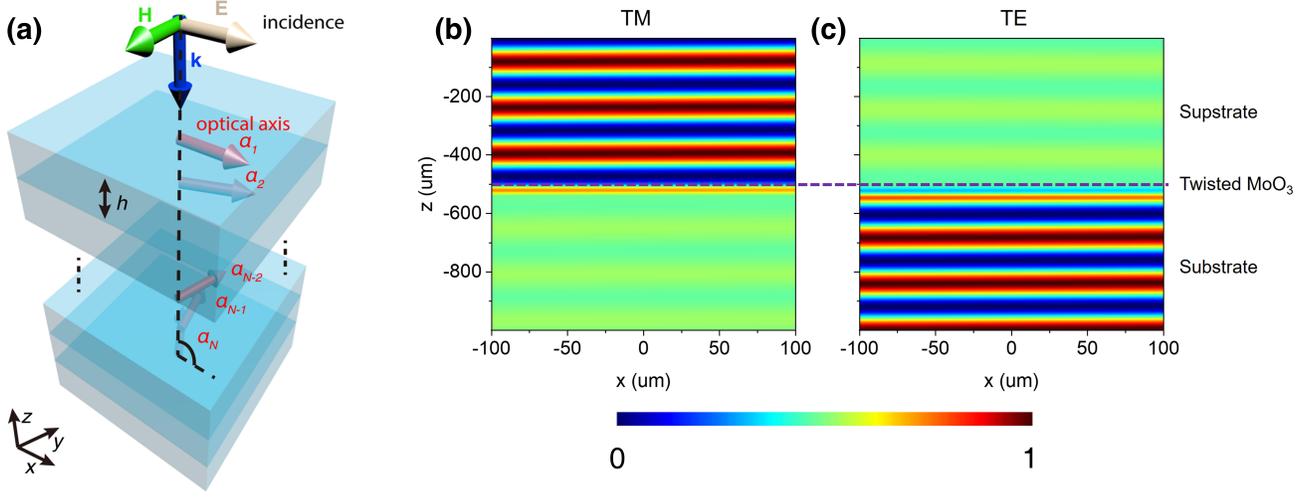

**Fig.1** Construction of the proposed tunable polarization rotator for midinfrared waves. (a) 3D view of the gradiently-twisted multilayers. (b-c) Electric field profiles of incident TM waves (b) and transmissive TE waves (c), respectively.

Without loss of generality, we consider a gradiently-twisted multiple α-MoO₃ layers as shown in **Fig.1a**. Here, the thickness of the α-MoO₃ thin-film is denoted as h; the total layer number is denoted as N; the twisted angle of each layer is ranging from $\alpha_1$ to $\alpha_N$ with $\alpha_n = n \times \alpha_N/N$. We theoretically calculate the optical responses of the gradiently-twisted multiple α-MoO₃ layers within the Maxwellian framework. The in-plane permittivities of a α-MoO₃ thin-film are determined from the Lorentz model, i.e.,

$$\varepsilon_{r1,2} = \varepsilon_{\infty 1,2}\left(1 + \frac{\omega_{LO1,2}^2 - \omega_{TO1,2}^2}{\omega_{LO1,2}^2 - \omega^2 - i\omega\Gamma}\right) \quad (1)$$

where the subscripts 1 and 2 represent the [100] and [001] directions of α-MoO₃. The adopted parameters for α-MoO₃ are taken from the experimental work [42]. Due to the subwavelength thickness, the α-MoO₃ thin-film is equivalent to a conductive surface, whose conductivity tensor is given by,

$$\overline{\overline{\sigma}}_{s,0} = \begin{bmatrix} \sigma_{s,\parallel} & 0 \\ 0 & \sigma_{s,\perp} \end{bmatrix} \quad (2)$$

where $\sigma_{s,\parallel} = -i\omega h \varepsilon_0 \varepsilon_{r1}$ and $\sigma_{s,\perp} = -i\omega h \varepsilon_0 \varepsilon_{r2}$.

For the convenience of analytical calculation, we introduce a global coordinate $(x, y, z)$ and a local coordinate $(\rho, h, z)$. For the local coordinate, the $\rho - z$ plane coincides the incident plane. In other words, the wavevector of the incidence can be defined as $\overline{k}_{in} = (q, 0, k_z)$ in the $(\rho, h, z)$ coordinate, where $q^2 + k_z^2 = (\omega/c)^2$ and the incident angle is $\theta = \mathrm{atan}(q/k_z)$. If one define the angle from the [100] direction to x-axis as $\alpha$ and the angle from the x-axis to the ρ-axis as $\kappa$, the surface conductivity in the $(\rho, h, z)$ coordinate is obtained as

$$\overline{\overline{\sigma}}_{s,\rho hz} = \overline{\overline{R}}_\kappa \overline{\overline{R}}_\alpha \overline{\overline{\sigma}}_{s,0} \overline{\overline{R}}_\alpha^{-1} \overline{\overline{R}}_\kappa^{-1} = \begin{bmatrix} \sigma_{\rho\rho} & \sigma_{\rho h} \\ \sigma_{h\rho} & \sigma_{hh} \end{bmatrix} \tag{3}$$

where $\overline{\overline{R}}_\alpha = \begin{bmatrix} \cos\alpha & \sin\alpha \\ -\sin\alpha & \cos\alpha \end{bmatrix}$, and $\overline{\overline{R}}_\kappa = \begin{bmatrix} \cos\kappa & -\sin\kappa \\ \sin\kappa & \cos\kappa \end{bmatrix}$.

On the other hand, the electromagnetic fields in an arbitrary region $j$ between conductive surfaces at $z = z_{j-1,j}$ and $z = z_{j,j+1}$ are composed of transverse-magnetic (TM) field and transverse-electric (TE) field. To be specific, $\overline{E} = \overline{E}_{TE} + \overline{E}_{TM}$ and $\overline{H} = \overline{H}_{TE} + \overline{H}_{TM}$. The TM field takes the form as followings:

$$\overline{E}_{TM} = qe^{iqu}\left(t^+_{m,j}\begin{bmatrix} -k_z \\ 0 \\ -q \end{bmatrix} e^{-ik_z(z-z_{j-1,j})} + r^-_{m,j}\begin{bmatrix} k_z \\ 0 \\ -q \end{bmatrix} e^{ik_z(z-z_{j,j+1})}\right) \tag{4}$$

$$\overline{H}_{TM} = \omega\varepsilon_0 qe^{iqu}\left(t^+_{m,j}\begin{bmatrix} 0 \\ 1 \\ 0 \end{bmatrix} e^{-ik_z(z-z_{j-1,j})} + r^-_{m,j}\begin{bmatrix} 0 \\ 1 \\ 0 \end{bmatrix} e^{ik_z(z-z_{j,j+1})}\right) \tag{5}$$

The TE field takes the form as followings:

$$\overline{E}_{TE} = k_0 qe^{iqu}\left(t^+_{e,j}\begin{bmatrix} 0 \\ 1 \\ 0 \end{bmatrix} e^{-ik_z(z-z_{j-1,j})} + r^-_{e,j}\begin{bmatrix} 0 \\ 1 \\ 0 \end{bmatrix} e^{ik_z(z-z_{j,j+1})}\right) \tag{6}$$

$$\overline{H}_{TE} = -\frac{k_0 qe^{iqu}}{\omega\mu_0}\left(t^+_{e,j}\begin{bmatrix} -k_z \\ 0 \\ -q \end{bmatrix} e^{-ik_z(z-z_{j-1,j})} + r^-_{e,j}\begin{bmatrix} k_z \\ 0 \\ -q \end{bmatrix} e^{ik_z(z-z_{j,j+1})}\right) \tag{7}$$

To determine the transmission coefficients $t^+_{m,e}$ and reflection coefficients $r^-_{m,e}$, we need to apply the following boundary conditions at each conductive surface:

$$\hat{n}_{j\to k} \times (\overline{E}_k - \overline{E}_j) = 0 \tag{8}$$

$$\hat{n}_{j\to k} \times (\overline{H}_k - \overline{H}_j) = \overline{\overline{\sigma}}_{s,\rho hz}\begin{bmatrix} E_{k,\rho} \\ E_{k,h} \end{bmatrix} \tag{9}$$

Following such a procedure, one can readily calculate the field distribution, transmission spectrum and reflection spectrum under arbitrary twisted angles and incident angles of the multilayer twisted α-MoO3 system (see more calculation details in Supplementary Material).

Our designed structure enables efficient conversion between TE and TM modes in a compact platform. To illustrate this, the field distributions of propagating modes are depicted in **Fig.1b** and **Fig.1c** when TM waves (19 THz) are used to illuminate the structure with the parameters: h=100nm, N=50, incident angle=1° and α$_N$=93.6°. The intensity of TM waves approaches to zero (**Fig.1b**) while TE waves appear (**Fig.1c**) after TM waves passing through the subwavelength multilayers. This suggests that the energy of TM waves is mostly converted into TE waves and the polarization of incident waves is rotated by 90°.

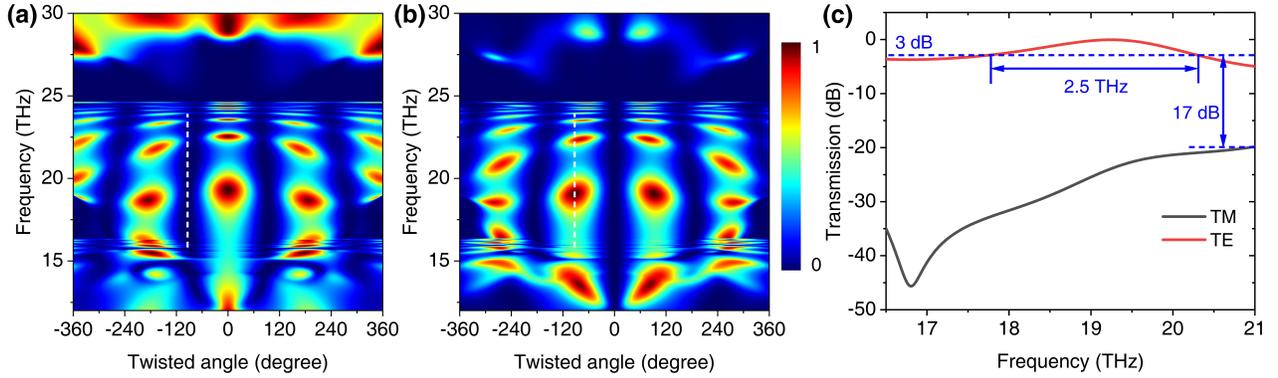

**Fig.2** Transmission of broadband linear polarization rotator illuminated by TM modes. (a-b) Transmission profile of TM mode (a) and TE mode (b) as a function of frequency and twisted angle. (c) The transmission spectrum of TM and TE with $\alpha_N = 93.6°$ (dashed lines in a and b) with gradiently-twisted structure parameters: *h*=100nm, *N*=50.

To further analyze the performance of gradiently-twisted MoO$_3$ based rotator, the transmission of the structure is calculated as a function of frequency and twisted angles. **Fig.2a** and **Fig.2b** show the transmission intensity contour plot of TM and TE when TM waves are used to illuminate the structure with *h*=100 nm and *N*=50. Both TM and TE demonstrate multiband transmission at certain twisted angle. At twist angle from 60° to 120°, TM waves disappears while the TE polarized waves appear with multiple bands after the light waves passing through the structure. **Fig.2c** shows the transmission plot of TM and TE waves with respect to frequency with $\alpha_N = 93.6°$. The TE polarized waves demonstrate a wide 3dB bandwidth of 2.5 THz with conversion efficiency near 100% at 19.03 THz (15.375 um), which demonstrates the capability of linear polarization rotation by 90° with high polarization ratio of 17 dB. This large bandwidth operation is related to the strong anisotropy of MoO$_3$ thin-film in a broad range of frequencies, while the high polarization ratio and conversion efficiency indicate nearly all energy of TM polarized waves is rotated and converted to TE polarized waves. The device performance demonstrates insensitivity to deviations in the number of layers and twisted angle, thereby facilitating practical device fabrication (see more details in Supplementary Material).

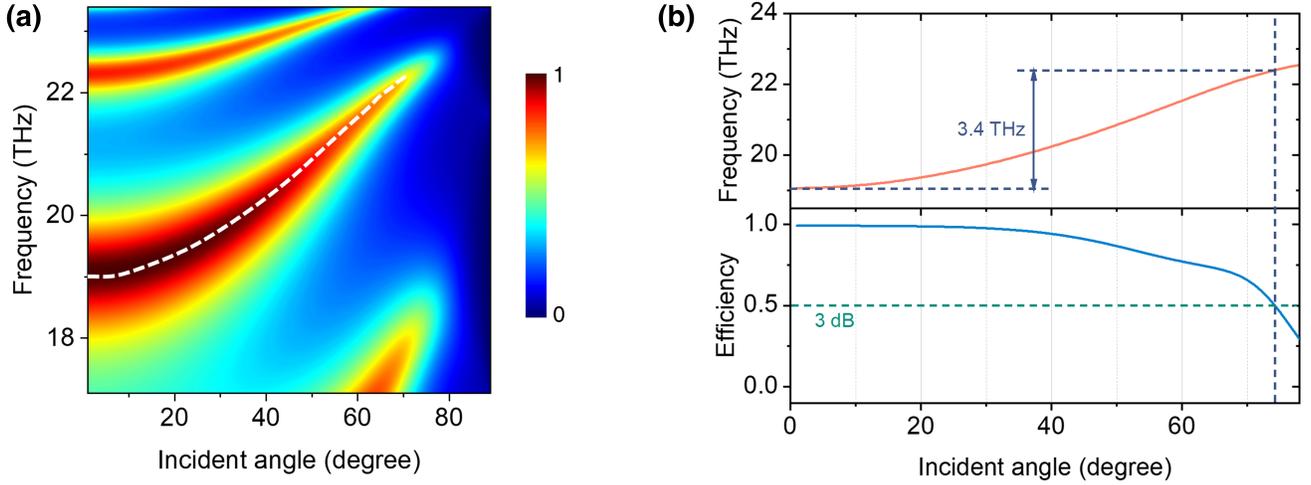

**Fig.3** Dependence of working frequency on the incident angles in gradiently-twisted structure with $\alpha_N = 93.6°$, $h$=100nm and $N$=50. (a) Transmission profile of TE mode as a function of frequency and incident angle. (b) Working frequency and efficiency as a function of incident angle.

To study the flexibility of the thin film rotator in manipulating operational frequency, we calculate the transmission spectrum of TE waves at various incident angles when TM waves illuminate the gradiently-twist structure. For simplicity, we choose the structure with the following parameters: $\alpha_N = 93.6°$, $h$=100nm and $N$=50. As shown in **Fig.3a**, the peak frequency (white dash line) of TE polarized waves shifts into higher frequency when the incident angles changes from 0 to 90°. The bandwidth experiences deterioration at high resonance frequencies in Fig.3a, attributed to the larger impendence mismatch when incident angle approaches 90° (see more details in Supplementary Material). To intuitively understand the working frequency tunability, **Fig.3b** plots the frequency and its corresponding efficiency with respect to the incident angle. The operational frequency is tunable from 19 Thz to 22.4 Thz at 3dB efficiency while the conversion efficiency gradually drops when increasing the incident angle, posing a trade-off between the tunability and efficiency owing to the strong impedance mismatch at large incident angle.

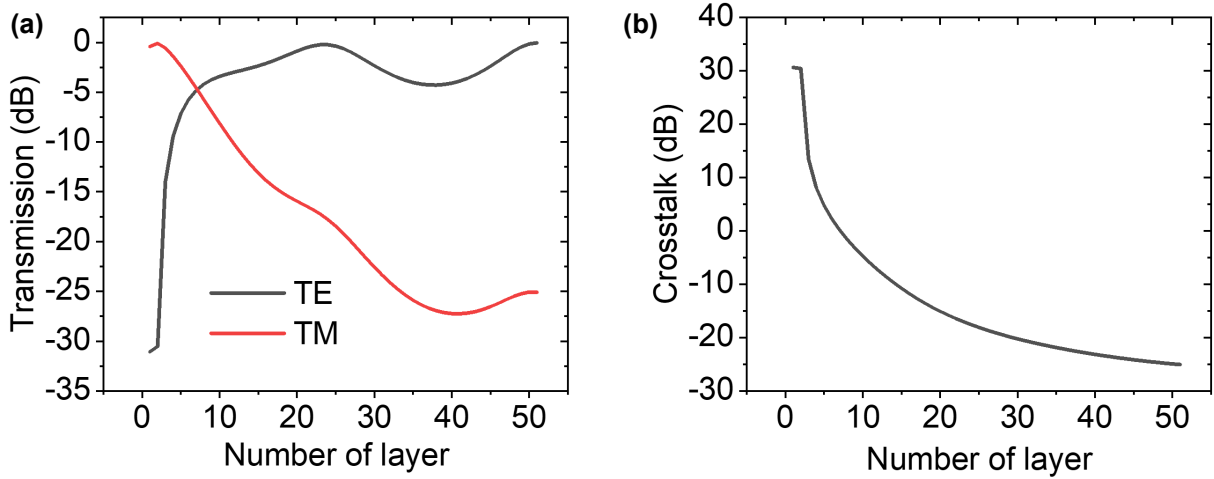

**Fig.4** Transmission of TE and TM as a function of layers. (a) Transmission of TE and TM as a function of layer number. (b) Crosstalk of the rotator as a function of layer number.

Finally, we also study the effect of layer number on the polarization rotation in a structure with h=100nm, $\alpha_N = 93.6°$ and incident angle=1°. As shown in **Fig.4a** and **Fig.4b**, increasing the number of layers enables more and more energy of incident waves rotated by 90°. Thus, the incident TM waves are converted to TE waves when passing through the structures and leading to smaller TM waves crosstalk at output.

In summary, we have theoretically studied a linear polarization rotator based on the gradiently-twisted α-MoO₃ multilayers. By twisted multilayers, the device can rotate the polarization of a TM waves by 90°with high conversion efficiency and low crosstalk. The proposed device has a subwavelength footprint with a thickness less than one-third of the operational wavelength. The gradiently-twisted structure exhibits the capability to engineer the optical activity while maintaining low optical losses, thereby enabling a wide bandwidth exceeding 2.5 THz with a polarization ratio surpassing 17dB. Moreover, the operational frequency is tunable across a 3.4 THz range by altering the incident angle, which gives the flexibility of the polarization rotator to freely tune the central wavelength. This high efficiency, large working bandwidth, high polarization ratio and freely tunable compact device provides more application potentials for free electron light sources[43-45], biosensing[46], wireless secure communication[2, 47] and imaging[48].

**Supplementary Material**

See the Supplementary Material for further information regarding the closed-form formulas of the design, as well as the impact of layer number and twisted angle on device performance.

**Acknowledgments**

The authors acknowledge funding support from Qinchuangyuan Program (Grant No. QCYRCXM-2023-095); Natural Science Basic Research Program of Shaanxi (Grant No. 2024JC-YBQN-0682); National Key R&D Program of China (Grant No. 2020YFB1807300); Distinguished Professor Fund of Jiangsu Province (Grant No. 1004-YQR23064); Selected Chinese Government Talent-recruitment Programs of Nanjing (Grant No. 1004-YQR23122); Startup Grant of Nanjing University of Aeronautics and Astronautics (Grant No. 1004-YQR23031).

**Data Availability**

The data that support the findings of this study are available from the corresponding author upon reasonable request.